\documentclass[pra,twocolumn,superscriptaddress]{revtex4-1}

\usepackage[utf8]{inputenc}
\usepackage[english]{babel}
\usepackage[T1]{fontenc}
\usepackage{lmodern}

\usepackage{graphicx}
\usepackage{amsmath}
\usepackage{amsfonts}
\usepackage{amssymb}
\usepackage{amsthm}

\usepackage{mathtools}

\usepackage{dsfont}
\usepackage{microtype}

\usepackage[breaklinks=true,colorlinks]{hyperref}

\newcommand{\cf}{cf.\ }

\newcommand{\coloneq}{\mathrel{\mathop:}=}

\newcommand{\dd}{\mathrm{d}}
\newcommand{\bd}{\begin{equation*}}
\newcommand{\ed}{\end{equation*}}

\newcommand{\Tr}{\operatorname{Tr}}

\newcommand{\ket}[1]{\left|{#1}\right\rangle}

\newcommand{\ketbra}[2]{\left|{#1}\middle\rangle\middle\langle{#2}\right|}
\newcommand{\proj}[1]{\ketbra{#1}{#1}}

\newcommand{\kB}{k_\mathrm{B}}
\newcommand{\indexc}{\mathrm{c}}
\newcommand{\indexh}{\mathrm{h}}

\newcommand{\betac}{\beta_\indexc}

\newcommand{\Tc}{T_\indexc}
\newcommand{\Th}{T_\indexh}

\newcommand{\sta}{\mathrm{STA}}
\newcommand{\cd}{\mathrm{CD}}
\newcommand{\ad}{\mathrm{ad}}
\newcommand{\na}{\mathrm{na}}
\newcommand{\Hcd}{H_\cd}
\newcommand{\Hsta}{H_\sta}
\newcommand{\Wad}{W_\ad}
\newcommand{\Wna}{W_\na}
\newcommand{\Wcd}{W_\cd}
\newcommand{\Wsta}{W_\sta}
\newcommand{\Qc}{Q_\indexc}
\newcommand{\Qh}{Q_\indexh}
\newcommand{\lambdac}{\lambda_\indexc}
\newcommand{\lambdah}{\lambda_\indexh}

\newcommand{\panel}[1]{\textcolor{red}{#1}}

\begin{document}

\title{Many-body quantum heat engines with shortcuts to adiabaticity}

\author{Andreas Hartmann}
\email{Andreas.Hartmann@uibk.ac.at}
\affiliation{Institut f\"ur Theoretische Physik, Universit\"at Innsbruck, Technikerstra{\ss}e~21a, A-6020~Innsbruck, Austria}
\author{Victor Mukherjee}
\email{mukherjeev@iiserbpr.ac.in}
\affiliation{Department of Physical Sciences, IISER Berhampur, Berhampur 760010, India}
\author{Wolfgang Niedenzu}
\email{Wolfgang.Niedenzu@uibk.ac.at}
\affiliation{Institut f\"ur Theoretische Physik, Universit\"at Innsbruck, Technikerstra{\ss}e~21a, A-6020~Innsbruck, Austria}
\author{Wolfgang Lechner}
\email{Wolfgang.Lechner@uibk.ac.at}
\affiliation{Institut f\"ur Theoretische Physik, Universit\"at Innsbruck, Technikerstra{\ss}e~21a, A-6020~Innsbruck, Austria}

\begin{abstract}
Quantum heat engines are modeled by thermodynamic cycles with quantum-mechanical working media. Since high engine efficiencies require adiabaticity, a major challenge is to yield a nonvanishing power output at finite cycle times. Shortcuts to adiabaticity using counter-diabatic (CD) driving may serve as a means to speed up such, otherwise infinitely long, cycles. We introduce local approximate CD protocols for many-body spin quantum heat engines and show that this method improves the efficiency and power for finite cycle times considerably. The protocol does not require \emph{a priori} knowledge of the system eigenstates and is thus realistic in experiments.
\end{abstract}

\date{\today}

\maketitle

\section{Introduction}

Heat engines are thermodynamic machines that cyclically convert heat into work~\cite{cengelbook}. Recently, the concept of heat engines has also been successfully applied in the quantum domain and constitutes an important and very active research direction within the emergent field of quantum thermodynamics, both theoretically~\cite{alicki1979quantum,kosloff1984quantum,kosloff2013quantum,gelbwaser2015thermodynamics,vinjanampathy2016quantum,kosloff2017quantum,binder2019thermodynamicsbook} and experimentally~\cite{koski2014experimental,rossnagel2016single,klaers2017squeezed,peterson2019experimental,vonlindenfels2019spin,klatzow2019experimental}. The quantum counterparts of, e.g., the Otto cycle, are considered to consist of a quantum system that is cyclically put into contact with two (hot and cold) heat baths and a work reservoir~\cite{kosloff2017quantum}. 

\par

A major difference between quantum and classical heat engines is the role of the adiabatic condition. Quantum-mechanically it does not suffice to implement the work-exchange strokes (devoid of any dissipative coupling to the environment) in an isentropic fashion to make the heat-to-work conversion as efficient as possible. Much rather, these strokes must be \emph{adiabatic in the quantum sense}. Quantum mechanically, a process is adiabatic if a system remains in its instantaneous eigenstate under an external change of the Hamiltonian, which requires the latter to be slow~\cite{born1928beweis,kato1950adiabatic,messiahbook,berry2009transitionless}. By contrast, fast changes would excite coherences in the system, i.e., the population of nondiagonal elements in its density matrix. Hence, while quantum mechanically an adiabatic process is always isentropic---as the system evolves in a unitary fashion according to the von~Neumann equation---the converse, however, is not true.

\par

A new strategy of overcoming the bottleneck of requiring adiabatically slow work-exchange strokes in quantum mechanical heat engines is to make use of so-called \emph{shortcuts to adiabaticity} (STA) methods~\cite{deng2013boosting, deffner2014classical, delcampo2014more, beau2016scaling, abah2017energy, campbell2017tradeoff, patra2017shortcuts, abah2018performance, diao2018shortcuts, duncan2018shortcuts, alipour2019shortcuts, abah2019shortcut, cakmak2019spin, delcampo2019focus, guery2019shortcuts, villazon2019swift}, which have also been applied experimentally~\cite{bason2012high,an2016shortcuts}. Therein, the initial slow adiabatic process is replaced by a different protocol (the shortcut) that ideally yields the same final state as the initial protocol---yet in finite time~\cite{torrontegui2013shortcuts}. In the context of quantum heat engines (QHEs), STAs have mostly been applied to single-body working media, e.g., a quantum harmonic oscillator~\cite{delcampo2014more,abah2018performance,abah2019shortcut,abah2019shortcutto,dupays2019shortcuts} or a single spin~\cite{cakmak2019spin}. Recently, for many-body systems local STA methods~\cite{mukherjee2016local,sels2017minimizing} and STAs across critical points~\cite{delcampo2012assisted} have been developed to efficiently speedup adiabatic protocols. It is thus a natural question whether these STA techniques can also be efficiently applied to many-body QHEs~\cite{campisi2016power,jaramillo2016quantum,chen2019interaction}.

\par

In this work we propose a four-stroke (two isentropic and two thermal) many-body quantum Otto engine that is sped up by shortcuts to adiabaticity to deliver finite power at finite speed. The isentropic strokes are driven using a recently developed approximate local counter-diabatic (CD) Hamiltonian following Refs.~\cite{sels2017minimizing} and \cite{hartmann2019rapid}. The engine's working medium is an Ising spin chain with nearest-neighbor interactions. We analytically derive the expressions for the CD terms and numerically simulate the engine for up to eight spins. These simulations reveal a considerable enhancement of the performance of these STA heat engines compared to their finite-time, and therefore nonadiabatic, analogs with respect to both efficiency and power. We stress that the derivation of the additional approximate counter-diabatic term \emph{does not require a priori knowledge of the system eigenstates} and can be implemented efficiently. We analyze the energetic balance of the cycle to gain further insight into the operational principles of the sped-up engine. Strikingly, we find that owing to the approximate nature of the STA the heat engine may be undesirably converted into a hybrid engine that is energized not only by heat but also by work stemming from the external control device that implements the CD drive.

\par

This paper is organized as follows. In Sec.~\ref{sec_qhe} we explain the major properties of many-body quantum heat engines, in particular, the quantum Otto cycle, and describe our models for the many-body working medium. In Sec.~\ref{sec_sta} we introduce shortcuts to adiabaticity using local counter-diabatic driving and its application to the quantum Otto cycle. We discuss the operational meaning of these STAs in Sec.~\ref{sec_opm} and numerically analyze the performance of the sped-up engine in Sec.~\ref{sec_pfn}. In Sec.~\ref{sec_dis} we conclude and give an outlook on future research.

\section{Quantum heat engine}\label{sec_qhe}

\subsection{Quantum Otto cycle}\label{sec_otto}

\begin{figure}
  \centering
  \includegraphics[width=.98\columnwidth]{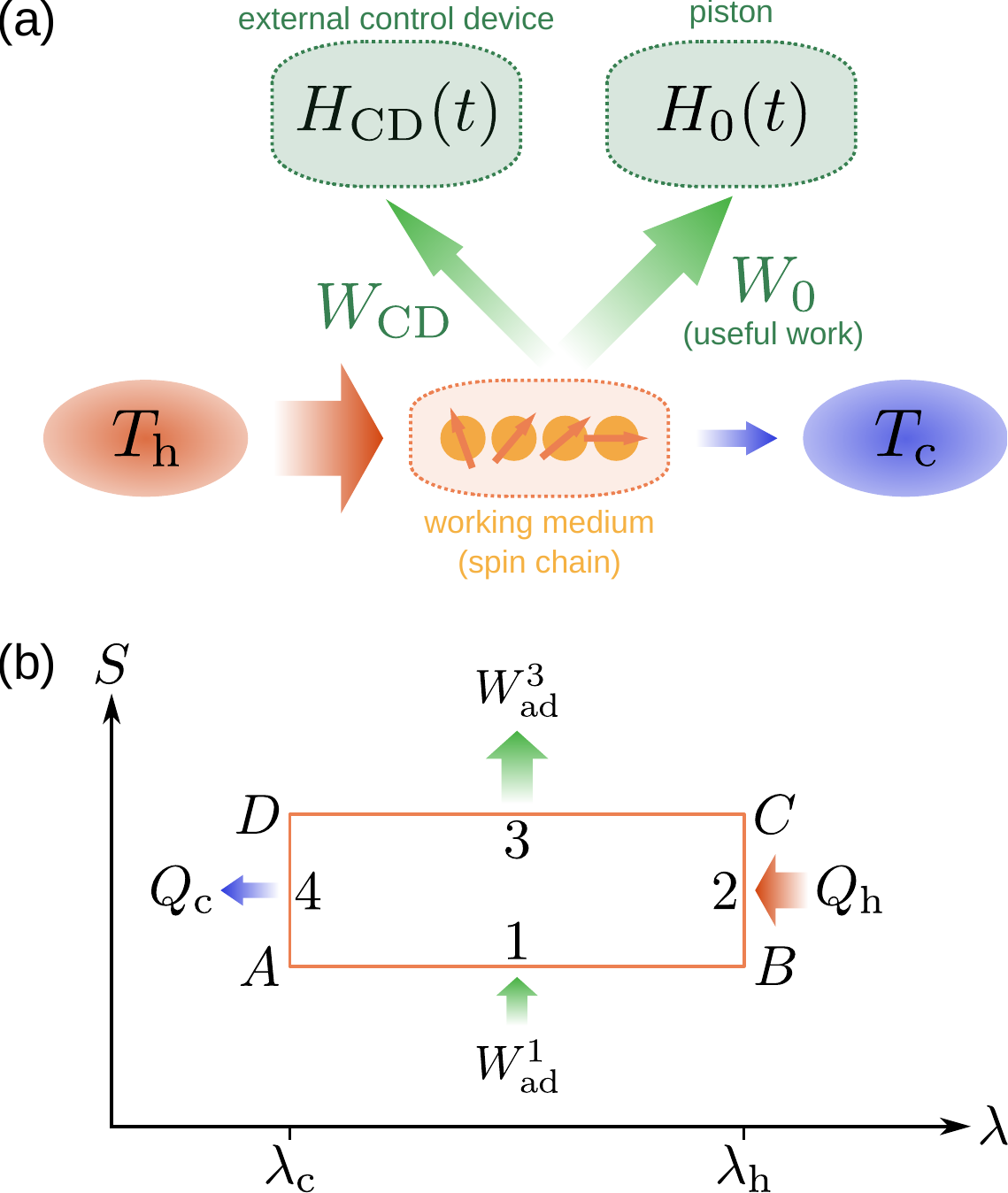}
  \caption{\textbf{Finite-time quantum heat engine} (a)~The working medium (spin chain) interacts with two thermal baths at temperatures $\Tc$ and $\Th$, respectively. Work to the load is extracted via the time-dependent protocol $H_0(t)$ implemented by a work reservoir (the load). In the underlying quantum Otto cycle this protocol must ideally be adiabatic to avoid the occurrence of ``quantum friction'' that would reduce the engine efficiency. The infinite cycle time required by the adiabaticity condition causes vanishing output power (work per cycle divided by the cycle time). A ``shortcut to adiabaticity'' is realized if an additional controller implements the right additional counter-diabatic protocol $\Hcd(t)$ on the working medium. The original (adiabatic) cycle is then significantly sped up such that the engine yields finite power. (b)~Adiabatic quantum Otto cycle in the $\lambda$--entropy plane. It consists of two unitary (hence isentropic) strokes ($1$ and $3$) with corresponding work $\Wad^1$ and $\Wad^3$ and two thermal strokes ($2$ and $4$) with corresponding heat $\Qc$ and $\Qh$, respectively. In the sped-up cycle the adiabatic protocol $H_0(t)$ is supported by an additional counter-diabatic drive $\Hcd(t)$ implemented by an external control device.}\label{fig:fig1}
\end{figure}

A four-stroke quantum Otto cycle~\cite{kosloff2017quantum} consists of two heat-exchange strokes, wherein the working medium (WM) is alternatingly coupled to two (hot and cold) thermal baths at temperatures $\Th$ and $\Tc$, respectively, and two work-exchange strokes. In the latter, the WM is isolated from the environments and its Hamiltonian $H_0(\lambda(t))$ is externally controlled via the time-dependent working parameter $\lambda(t)$. To operate as an engine (produce work), the adiabatic Otto cycle is traversed in the following order [see Fig.~\ref{fig:fig1}\panel{(b)}].
\begin{enumerate}
\item \emph{Stroke 1: Adiabatic compression} ($A \rightarrow B$). Initially, the WM is in the thermal state $\rho_A=e^{-\beta_{\mathrm{c}} H_0(\lambda_\indexc)}/Z(\lambda_\indexc)$ at inverse temperature $\betac=1/(\kB\Tc)$ and Hamiltonian $H_0(\lambda_\indexc)$, where $\lambda_\indexc  \coloneq \lambda(t=0)$ and $Z(\lambda_\indexc)=\Tr[e^{-\beta_{\mathrm{c}} H_0(\lambda_\indexc)}]$ is the partition function. The working parameter $\lambda(t)$ is adiabatically increased from $\lambda_\indexc$ to $\lambda_\indexh \coloneq \lambda(t=\tau_1)$ such that the populations of the instantaneous eigenstates of $H_0(\lambda(t))$ remain invariant. At the end of the stroke the WM attains the state $\rho_B$. Hence, the work
  \begin{subequations}\label{eq_energies_ad_otto}
  \begin{equation}
    \Wad^1 \coloneq \langle H_0(\lambda_\indexh) \rangle_{\rho_B}  - \langle H_0(\lambda_\indexc) \rangle_{\rho_A}  
  \end{equation}
  is performed on the WM.
\item \emph{Stroke 2: Hot isochore} ($B \rightarrow C$). The WM is brought into contact with the hot thermal bath while its Hamiltonian $H_0(\lambda_\indexh)$ remains constant. The stroke time $\tau_2$ is sufficiently long such that the WM thermalizes to the state $\rho_C=e^{-\beta_{\mathrm{h}} H_0(\lambda_\indexh)}/Z(\lambda_\indexh)$. During this stroke, the heat
  \begin{equation}
    Q_\indexh \coloneq \langle H_0(\lambda_\indexh) \rangle_{\rho_C}  - \langle H_0(\lambda_\indexh) \rangle_{\rho_B}
  \end{equation}
  is imparted by the hot bath.
\item \emph{Stroke 3: Adiabatic expansion} ($C \rightarrow D$). The working parameter decreases adiabatically from $\lambda_\indexh$ to $\lambda_\indexc$ in the stroke time $\tau_3$ and the WM attains the state $\rho_D$. Hence, the work
  \begin{equation}
    \Wad^3 \coloneq \langle H_0(\lambda_\indexc) \rangle_{\rho_D}  - \langle H_0(\lambda_\indexh) \rangle_{\rho_C}
  \end{equation}
  is extracted from the WM.
\item \emph{Stroke 4: Cold isochore} ($D \rightarrow A$). The WM is brought into contact with the cold thermal bath, where for a sufficiently long stroke time $\tau_4$ the WM cools back down into the initial state $\rho_A$ of the cycle. Hence, the heat
  \begin{equation}
    Q_\indexc \coloneq \langle H_0(\lambda_\indexc) \rangle_{\rho_A}  - \langle H_0(\lambda_\indexc) \rangle_{\rho_D}
  \end{equation}
  \end{subequations}
  is transferred to the cold bath.
\end{enumerate}
Note that engine operation (work extraction) corresponds to $\Wad^{1+3}\coloneq\Wad^1 + \Wad^3 < 0$. 
Here we use the expressions ``compression'' and ``expansion'' in analogy with the classical Otto engine where $\Wad^1 >0$ and $\Wad^3 <0$. Depending on the physical implementation, however, the roles of the two strokes (i.e., the signs of $\Wad^1$ and $\Wad^3$) may be interchanged~\cite{geva1992quantum}. 

The efficiency of this adiabatic cycle is the net work performed by the WM on the piston divided by the heat transferred from the hot bath to the WM, i.e.,
\begin{equation}
\eta = - \dfrac{\Wad^1 + \Wad^3}{Q_{\mathrm{h}}},
\end{equation}
and is limited by the Carnot efficiency, $\eta \leq 1-\Tc/\Th$.

\par

The power of the engine is given by the work done by the WM divided by the total cycle time $\tau_{\mathrm{cycle}}=\sum_{l=1}^4 \tau_l$. If the Hamiltonian does not commute with itself at all times, $[H_0(t),H_0(t^\prime)]=0\ \forall t,t^\prime$, the adiabaticity condition requires infinitely long durations of strokes $1$ and $3$, $\tau_1,\tau_3\rightarrow\infty$. Consequently, in the adiabatic limit the power
\begin{equation}
  \mathcal{P}_\ad=\lim_{\tau_1,\tau_3\rightarrow\infty}\dfrac{\Wad^1 + \Wad^3}{\tau_{\mathrm{cycle}}}\rightarrow 0
\end{equation}
vanishes, which renders the engine practically useless.

\par

The first way to circumvent this issue would be to apply the protocol $H_0(t)$ in a finite time, thus giving up the strict requirement for adiabatic compression or expansion. The price of these nonadiabatic dynamics is the occurrence of so-called ``quantum friction'' (excitation of coherences)~\cite{kosloff2002discrete,feldmann2003quantum,feldmann2006quantum}, which reduces the output work. Being traversed in a finite time, this nonadiabatic engine yields finite power (recall that a negative sign indicates power output), 
\begin{equation}
  \mathcal{P}_\na=\dfrac{\Wna^1 + \Wna^3}{\tau_{\mathrm{cycle}}}<0.
\end{equation}
This approach, however, has two caveats: (i)~Quantum friction may significantly reduce the engine efficiency (as the work per cycle is reduced), and (ii)~for too short cycle times the machine may cease to act as an engine, $P_\na>0$, due to $\Wna^{1+3}\coloneq \Wna^1+\Wna^3$ becoming positive, which corresponds to \emph{work consumption} rather than \emph{work extraction}.

\par

A possible solution of this dilemma is to introduce an external contol device that applies an additional counter-diabatic drive $\Hcd(t)$ to the working medium [see Fig.~\ref{fig:fig1}\panel{(a)}]. The resulting protocol $H_0(t)+\Hcd(t)$ is known as a \emph{shortcut to adiabaticity} (STA)~\cite{deng2013boosting,delcampo2014more,beau2016scaling,abah2017energy,campbell2017tradeoff,abah2018performance,abah2019shortcut,cakmak2019spin,  alipour2019shortcuts, duncan2018shortcuts, diao2018shortcuts, delcampo2019focus,villazon2019swift}: It allows performance of the transformation $(\rho_A,H(\lambda_\indexc))\mapsto(\rho_B,H(\lambda_\indexh))$ (and similarly for the third stroke) in \emph{finite} time. This way the work per cycle equals its adiabatic counterpart but the cycle time is finite. Hence, the engine yields finite power while maintaining the ideal adiabatic efficiency.

\par

\emph{Exact} counter-diabatic protocols can be analytically derived in special cases, e.g., for a harmonic oscillator, a single spin, or two interacting spins~\cite{chen2011lewis, takahashi2013transitionless, delcampo2014more, abah2018performance, abah2019shortcut, abah2019shortcutto, dupays2019shortcuts, cakmak2019spin}. By contrast, for a many-body working medium (see Appendix~\ref{App:secA1} for the single-body case) we have to rely on an \emph{approximate} counter-diabatic drive. This, as we see, entails important operational consequences for the heat-engine operation.

\subsection{Many-body quantum working medium}\label{SecIIb}

We consider an Ising spin chain with nearest-neighbor interactions and Hamiltonian
\begin{equation}
H_0(t)=- \sum_{j=1}^{N} h_j(t) \sigma_j^x - \sum_{j=1}^{N} b_j(t) \sigma_j^z - \sum_{j=1}^{N} J_j(t) \sigma_j^z \sigma_{j+1}^z,
\label{eq:IsingSpinModel}
\end{equation}
where $N$ is the total number of spins, $h_j(t)$ and $b_j(t)$ are the time-dependent strengths of the magnetic fields at site $j$ in the $x$ and $z$ directions, respectively, and $J_j(t)$ is the time-dependent strength of the interaction between spins at sites $j$ and $j+1$, where we impose periodic boundary conditions, i.e., $\sigma_{N+1}=\sigma_{1}$. Recent progress in controlling many-body quantum systems has made it possible to experimentally realize and study similar many-body Hamiltonians using quasi one-dimensional Ising ferromagnets~\cite{brooke1999quantum,coldea2010quantum} or cold atoms~\cite{bernien2017probing}.

\par

The abstract working parameters $\lambdac$ and $\lambdah$ in the transferred energies, Eq.~\eqref{eq_energies_ad_otto}, then correspond to $h_j(t=0)=h_{j,\mathrm{i}}$, $b_j(t=0)=b_{j,\mathrm{i}}$, $J_j(t=0)=J_{j, \mathrm{i}}$ and $h_j(t=\tau_1)=h_{j,\mathrm{f}}$, $b_j(t=\tau_1)=b_{j,\mathrm{f}}$, $J_j(t=\tau_1)=J_{j, \mathrm{f}}$, respectively. The explicit forms of the magnetic fields $h_j(t)$ and $b_j(t)$ and the interactions $J_j(t)$ are given in Eqs.~\eqref{app_fields_many-body} in Appendix~\ref{app_sta_protocols}. 

\section{Shortcuts to adiabaticity}\label{sec_sta}

The adiabatic theorem~\cite{born1928beweis,kato1950adiabatic,messiahbook} poses a speed limit on quantum adiabatic processes. The precise role of the adiabatic condition in many-body quantum systems has recently regained interest with the emergence of adiabatic quantum computing~\cite{albash2018adiabatic, hauke2019perspectives}. A variety of STA methods~\cite{demirplak2003adiabatic, berry2009transitionless, chen2010fast, chen2011lewis, takahashi2013transitionless, jarzynski2013generating} including counter-diabatic driving \cite{delcampo2013shortcuts, damski2014counterdiabatic, sels2017minimizing} have been developed and applied in the field of adiabatic quantum computation~\cite{farhi2000quantum, farhi2001quantum, childs2001robustness} and quantum annealing~\cite{finnila1994quantum, kadowaki1998quantum, brooke1999quantum, santoro2002theory, santoro2006optimization, boixo2013experimental, boixo2014evidence}. Recently, it has also been shown experimentally that applying STA methods can dramatically enhance the performance of quantum annealing~\cite{an2016shortcuts}. 

\par

STAs have been successfully applied to QHEs with a single-body working medium~\cite{delcampo2014more,abah2018performance,abah2019shortcut,abah2019shortcutto,dupays2019shortcuts,cakmak2019spin}. It is thus a natural question whether they can also be applied to many-body quantum heat engines, where we have to rely on approximate solutions for the counter-diabatic drive. We note that we apply shortcuts only on the originally adiabatic strokes, as these are typically much slower than the thermalization strokes. Techniques such as shortcuts to equilibration for speeding up the dynamics of open quantum systems have recently been proposed~\cite{alipour2019shortcuts,dann2019fast,dann2019shortcut,dupays2019shortcuts}.

\subsection{Approximate counter-diabatic driving}\label{SecIIa}

For finite times, the original protocol $H_0(t)$ [here, Eq.~\eqref{eq:IsingSpinModel}] induces a nonadiabatic (diabatic) evolution by generating coherences in the working medium (``quantum friction'').
To avoid these coherences, shortcuts to adiabaticity are realized by evolving the working medium according to the Hamiltonian
\begin{equation}
H_{\mathrm{STA}}(t)=H_0(t)+H_{\mathrm{CD}}(t)
\label{eq:1}
\end{equation}
rather than only $H_0(t)$. The additional counter-diabatic Hamiltonian $H_\mathrm{CD}(t)$ compensates those undesirable nonadiabatic effects~\cite{torrontegui2013shortcuts}. The determination of its \emph{exact} form requires \emph{a priori knowledge} of the system eigenstates for all times, which, in the case of complex many-body working media, is impracticable for both numerical computations and experimental implementations (see Appendix~\ref{sec:AppendixB}).

\par

With this challenge in mind we resort to a recently proposed variational method for finding the counter-diabatic Hamiltonian~\cite{sels2017minimizing,kolodrubetz2017geometry}
\begin{equation}
\Hcd(t)=\dot{\vartheta}(t) \mathcal{A}_{\vartheta}(t), \label{eq:eq2}
\end{equation}
where $\mathcal{A}_{\vartheta}(t)$ is the so-called adiabatic gauge potential and $\vartheta(t)$ a control function. The goal is to find an \emph{approximate} expression for the CD protocol $H^*_\mathrm{CD}(t)$ by making a local ansatz $\mathcal{A}^*_{\vartheta}(t)$ that approximates the solution of $[i \partial_{\vartheta} H_0(t)-[\mathcal{A}^*_{\vartheta}(t),H_0(t)], H_0(t)]=0$ (see Appendix~\ref{sec:AppendixB}).

\subsection{Otto cycle with counter-diabatic driving}\label{SecIIc}

For the original Hamiltonian, Eq.~\eqref{eq:IsingSpinModel}, we use the \emph{local} ansatz
\begin{equation}
  \mathcal{A}_t^*(t) \coloneq \sum_{j=1}^N \alpha_j(t) \sigma_j^y \label{eq:ansatz1}
\end{equation}
to approximate the counter-diabatic Hamiltonian, Eq.~\eqref{eq:eq2}, and where $\mathcal{A}^*_t(t) = \dot{\vartheta}(t) \mathcal{A}^*_\vartheta(t)$ is the adiabatic gauge potential with respect to time $t$. This ansatz consists of applying additional magnetic fields in the $y$ direction for each spin. As shown in Appendix~\ref{sec:AppendixB1} the optimal solution for these fields evaluates to
\begin{equation}
  \alpha_j(t)=\dfrac{1}{2}\dfrac{\dot{h}_j(t) b_j(t)-\dot{b}_j(t) h_j(t)}{h_j(t)^2+b_j(t)^2+J_{j-1}(t)^2 + J_{j}(t)^2},
\label{eq:alphaspin}
\end{equation}
and thus the local shortcut-to-adiabaticity Hamitonian, Eq.~\eqref{eq:1}, adopts the form
\begin{multline}
H^*_{\mathrm{STA}}(t) = -\sum_{j=1}^{N} h_j(t) \sigma_j^x - \sum_{j=1}^{N} b_j(t) \sigma_j^z \\
 - \sum_{j=1}^{N} J_j(t) \sigma_j^z \sigma_{j+1}^z+\sum_{j=1}^{N} Y_j(\vartheta_0, t) \sigma_j^y,
\label{eq:SpinCDHamiltonian}
\end{multline}
where the asterisk denotes that the Hamiltonian is \emph{inexact}.
Here we have defined $Y_j(\vartheta_0, t)\coloneq\alpha_j(t) \dot{\vartheta}(\vartheta_0,t)$ with the control function
\begin{equation}
  \vartheta(\vartheta_0,t)\coloneq\vartheta_0  \sin^2\left[\dfrac{\pi}{2}\sin^2\left(\dfrac{\pi t}{2 \tau}\right)\right]
\label{eq:lambda}
\end{equation}
which assures smoothness at the beginning and end of the strokes (see Appendix~\ref{app_sta_protocols} for more information). Since
\begin{equation}\label{eq_Hcd}
  H^*_\mathrm{CD}(t) \coloneq \sum_{j=1}^{N} Y_j(\vartheta_0, t) \sigma_j^y
\end{equation}
is an inexact, approximate counter-diabatic drive, the resulting states at points $B$ and $D$ in Fig.~\ref{fig:fig1} will not exactly be $\rho_B$ and $\rho_D$, respectively, but different states, $\rho_B'$ and $\rho_D'$, with the same entropy but different energy. The reliability of $H^*_\mathrm{CD}(t)$, i.e., how well the target state is reached, can be greatly improved by a variation of the global strength parameter $\vartheta_0$ in Eq.~\eqref{eq:lambda}~\cite{hartmann2019rapid}.

\section{Work under STA protocols}\label{sec_opm}

During the unitary strokes [strokes $1$ and $3$ in Fig.~\ref{fig:fig1}\panel{(b)}] the dynamics of the working medium is governed by the time-dependent, Hamiltonian $\Hsta(t)$. Consequently, the energy change of the WM corresponds to the total exchanged work~\cite{pusz1978passive,lenard1978thermodynamical,alicki1979quantum}, 
\begin{equation}\label{eq_Wsta}
  \Wsta^j \equiv \Delta E = \int_0^{\tau_j} \Tr \left[\rho(t) \dot{H}_\sta(t)\right] \dd t,
\end{equation}
where $j \in \{1,3\}$ denotes the corresponding isentropic stroke and $\Hsta(t)=H_0(t)+\Hcd(t)$ [Eq.~\eqref{eq:1}] is the time-dependent shortcut-to-adiabaticity Hamiltonian. If $H_0(t)$ and the counter-diabatic drive $\Hcd(t)$ are implemented by two independent work reservoirs, the division of the total work, Eq.~\eqref{eq_Wsta}, into the two components,
\begin{subequations}\label{eq_W0_Wcd}
  \begin{align}
    W_0^j &\coloneq \int_0^{\tau_j} \Tr \left[\rho(t) \dot{H}_0(t)\right] \dd t, \label{eq_W0}\\
    \Wcd^j &\coloneq \int_0^{\tau_j} \Tr \left[\rho(t) \dot{H}_\cd(t)\right] \dd t, \label{eq_Wcd}
  \end{align}
\end{subequations}
is operationally interpreted as the individual work components exchanged between the working medium (the spin chain) and the two work reservoirs. Physically, this would correspond to the situation where the additional field in the $y$ direction in Eq.~\eqref{eq:SpinCDHamiltonian} is implemented by a second control unit (the external ``controller''), independent of the one that implements the original protocol (the piston or ``load'') [\cf Fig.~\ref{fig:fig1}\panel{(a)}]. Note, however, that in general $W_0$ is \emph{not} the same work as in the original adiabatic Otto cycle since $\rho(t)$ is now determined by $\Hsta(t)$ rather than $H_0(t)$. As we see below, this crucially depends on how good the chosen, \emph{approximate} STA protocol, Eq.~\eqref{eq:SpinCDHamiltonian}, reproduces the ideal, \emph{exact} STA protocol.

\par

In order to understand the consequences of the division, Eq.~\eqref{eq_W0_Wcd}, on the operation of the Otto engine we must distinguish between the cases where over a cycle $\Wcd^{1+3}\coloneq\Wcd^1+\Wcd^3\leq 0$ (heat engine) and $\Wcd^{1+3}> 0$ (thermomechanical engine). Irrespective of $\Wcd^{1+3}$, the machine operates as an engine (produces useful work) only if $W_0^{1+3}\coloneq W_0^1+W_0^3< 0$.

\subsection{Heat-engine regime}\label{sec_he}

If over a cycle $\Wcd^{1+3}\leq0$, the machine works as a genuine heat engine that converts thermal energy into work. This work, however, is performed not only on the work reservoir that implements $H_0(t)$ but also \emph{on the work reservoir that implements $\Hcd(t)$}. Hence, not all the work performed by the engine is available to the load. Hence, the \emph{useful} power generated by the engine is
\begin{equation}\label{eq_P_heat}
  \mathcal{P}\coloneq \frac{\mathrm{useful\ work\ output}}{\mathrm{cycle\ time}} = \frac{W_0^{1+3}}{\tau_\mathrm{cycle}}
\end{equation}
since $\Wcd$ is not available to the load. Consequently, the engine efficiency, as experienced by the load, is
\begin{equation}\label{eq_eta_heat}
 \eta_\mathrm{\hspace{1pt}heat} \coloneq \frac{\mathrm{useful\ work\ output}}{\mathrm{energy\ input}} \equiv \frac{-W_0^{1+3}}{\Qh}.
\end{equation}
Note that in this regime (where $\Wcd^{1+3} < 0$) the finite-time engine is solely energized by the heat $\Qh > 0$ stemming from the hot bath, which characterizes a genuine heat engine. This heat input is converted into the useful mechanical work output $W_0^{1+3} < 0 $. 

\par

Our numerical simulations (Sec.~\ref{sec_pfn}) show that the engine performing work on the control device is an \emph{artifact} of inexact counter-diabatic driving, i.e., when the final state generated by $\Hsta(t)$ in either stroke $1$ or stroke $3$ differs from the final state generated by the original protocol in the respective stroke: We currently do not have a general analytic proof but in our numerical simulations we could clearly observe that $\Wcd^{1+3}=0$ if the counter-diabatic term is exact (see the discussion of the single-body working medium in Appendix~\ref{App:secA1}), meaning that then $W_0^{1+3}$ \emph{equals} its counterpart $\Wad^{1+3}$ in the adiabatic Otto cycle in Sec.~\ref{sec_otto}. This observation clearly demonstrates the importance of striving for a perfect counter-diabatic protocol to speed up the Otto cycle. For a many-body working medium, however, the exact protocol is typically hard to find analytically and, even if it is known, may be extremely challenging to implement in an actual experiment since it will be of a nonlocal nature. Namely, it will not only involve single-body terms as in Eq.~\eqref{eq:SpinCDHamiltonian} but higher-order terms, possibly up to complicated $N$-body interactions.

\par

On the other hand, this feature enables us to optimize (to some extent) the counter-diabatic drive by trying to minimize $|\Wcd^{1+3}|$ experimentally.

\subsection{Hybrid thermomechanical engine regime}\label{sec_hybrid}

If the counter-diabatic protocol is not exact, we may also encounter situations in which $\Wcd^{1+3}>0$ over a cycle, which has striking operational consequences for the engine: Rather than being a genuine heat engine that converts thermal energy into useful work, the machine now acts as a \emph{hybrid thermomechanical engine}~\cite{niedenzu2016operation,dag2016multiatom,ghosh2017catalysis,niedenzu2018quantum} that is powered by thermal energy $\Qh > 0$ \emph{and an external battery} that provides $\Wcd^{1+3} > 0$. Nominally, $W_0^{1+3}$ may now strongly surpass its adiabatic counterpart but this work does not solely stem from converted thermal energy. Such sped-up engines could be compared to QHEs powered by non-thermal baths, e.g., squeezed-thermal baths, which are hybrid engines and as such are not bounded by the Carnot efficiency~\cite{niedenzu2016operation,dag2016multiatom,niedenzu2018quantum}. Naturally, despite its increased output power, speeding up a heat engine to the price of rendering it thermomechanically can be undesirable.

\par

Whereas the power of such a hybrid engine is still given by Eq.~\eqref{eq_P_heat}, its efficiency differs from its heat-engine counterpart, Eq.~\eqref{eq_eta_heat}, and reads
\begin{equation}\label{eq_eta_hybrid}
  \eta_\mathrm{\hspace{1pt}hybrid} \coloneq \frac{\mathrm{useful\ work\ output}}{\mathrm{energy\ input}} \equiv \frac{-W_0^{1+3}}{\Qh+\Wcd^{1+3}}.
\end{equation}
Note that in this regime (where $\Wcd^{1+3} > 0$) the finite-time engine is energized by the heat $\Qh > 0$ stemming from the hot bath as well as by the work $\Wcd^{1+3} > 0$ stemming from the external controller. This characterizes a hybrid thermomechanical engine. 
The combined heat and work input is converted into useful mechanical work output $W_0^{1+3} < 0$. While the power of this hybrid engine formally appears to be the same as for the heat engine [Eq.~\eqref{eq_P_heat}], the physical origin of $W_0^{1+3}$ strongly differs and its magnitude may strongly surpass its counterpart from the adiabatic Otto engine.

\par

We note that the above considerations only apply to the case where the working medium interacts with \emph{two independent} work reservoirs. If the counter-diabatic protocol $\Hcd(t)$ is also implemented by the piston, the controller ceases to be an external resource. The division, Eq.~\eqref{eq_W0_Wcd}, then becomes operationally irrelevant (even unmeasurable) and the machine operates as a genuine heat engine in either case with the total useful work output $\Wsta^{1+3} < 0$ and the energy input $\Qh > 0$. While in the nominal heat-engine regime $|\Wsta^{1+3}|>|W_0^{1+3}|$, in the nominal hybrid regime $|\Wsta^{1+3}|<|W_0^{1+3}|$ (we call these regimes ``nominal'' in the single-work-reservoir setup as they do not have an operational meaning). We note, however, that in most experimental setups the controller and load being two independent work reservoirs is probably the more natural situation.

\par

Finally, despite being detrimental to the engine operation if negative, we stress that the notion of $\Wcd^{1+3}$ being the work exchanged between the working medium and the controller strongly differs from other cost quantifiers discussed in the literature~\cite{abah2017energy,campbell2017tradeoff,zheng2016cost, abah2019shortcut, abah2018performance,cakmak2019spin,tobalina2019vanishing, abah2019shortcutto}: First, these costs pertain to exact protocols (where $\Wcd^{1+3}=0$), and, second, they quantify the extra energy that is required to \emph{implement} $\Hcd(t)$ for a certain time, e.g., the required intensity of an electric field. These costs of course become larger the shorter the unitary strokes become (as the fields become stronger and stronger). Since these costs are very strongly implementation-dependent we do not discuss them further in this paper, but note that they may significantly reduce the efficiency of sped-up engines~\cite{tobalina2019vanishing}.

\section{Numerical Performance Analysis}\label{sec_pfn}

\begin{figure*}
  \centering
  \includegraphics[width=\textwidth]{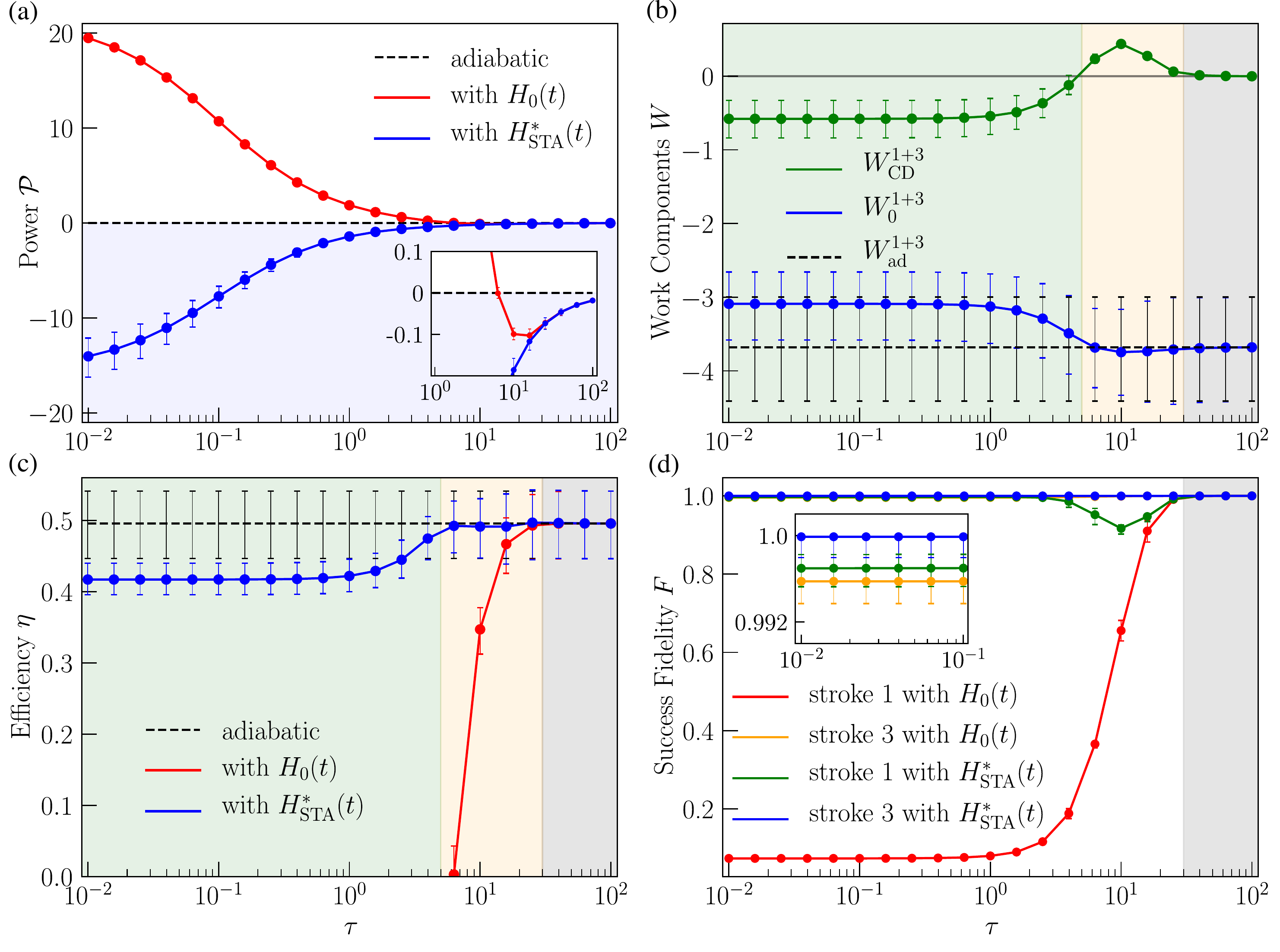}
  \caption{\textbf{Numerical simulation of the finite-time quantum heat engine with a many-body working medium.}
    (a)~Power $\mathcal{P}$ [Eq.~\eqref{eq_P_heat}] of the sped-up Otto cycle governed by (i)~the original protocol $H_0(t)$ [Eq.~\eqref{eq:IsingSpinModel}] and (ii)~the shortcut-to-adiabaticity protocol $H^*_\mathrm{STA}(t)$ [Eq.~\eqref{eq:SpinCDHamiltonian}] as a function of the isentropic-stroke duration $\tau=\tau_1=\tau_3$. The machine acts as an engine if $\mathcal{P}<0$ (blue-shaded area). Inset: Zoom-in on the region where the original protocol ceases to describe an engine for shorter cycle times. (b)~Work components $W_0^{1+3}$ and $W_\cd^{1+3}$ pertaining to the piston (load) and the external control device, respectively (\cf Fig.~\ref{fig:fig1}). The green (left)- and yellow (middle)-shaded areas depict the regions where the machine operates as a heat engine ($W_\cd^{1+3}<0$) and a thermomechanical engine ($W_\cd^{1+3}>0$), respectively. The gray (right)-shaded area depicts the adiabatic-limit region where $\Wcd^{1+3} < 10^{-3}$. (c)~Efficiency $\eta$ [Eq.~\eqref{eq_eta_heat} for the heat-engine regime (green-shaded area) and Eq.~\eqref{eq_eta_hybrid} for the hybrid thermomechanical regime (yellow-shaded area)]. (d)~Success fidelities [Eq.~\eqref{eq:Fidelity}] of the isentropic strokes with and without the STA protocol. Inset: Zoom-in. Parameters: duration of the isentropic strokes, $\tau_1=\tau_3=\tau$; duration of the thermalization strokes, $\tau_2=\tau_4=0.1$. The other parameters are $\Tc=0.22$, $\Th=22$, $h_{j,\mathrm{i}}=0.5$, $b_{j,\mathrm{i}}=0$, $h_{j,\mathrm{f}}=0$, $b_{z,\mathrm{f}}=1$ and $J_{j,\mathrm{i}}=0$ for each spin. We introduce disorder into the interaction strengths, where the $100$ final interaction strengths $J_{j,\mathrm{f}}$ are randomly chosen from a Gaussian distribution with standard deviation $\sigma=0.1$ and zero mean. The counter-diabatic drive is optimized by a control parameter $\vartheta_0$ bounded in $[0,1]$ (see text). Vertical bars denote the largest and lowest values of the power, work, efficiency, and success fidelity, respectively.}
  \label{fig:EfficienciesPowerCounter}
\end{figure*}

\par

As the main part of this work, we are interested in the performance of the sped-up many-body quantum Otto cycle. To this end we numerically compare the performance of Otto engines with [$H^*_\mathrm{STA}(t)$; Eq.~\eqref{eq:SpinCDHamiltonian}] and without [$H_0(t)$; Eq.~\eqref{eq:IsingSpinModel}] the counter-diabatic drive $H^*_\mathrm{CD}(t)$ for a system size of $N=8$ spins. Namely, we numerically integrate the von~Neumann equations $i \dot{\rho}_\mathrm{STA}(t)=[H^*_\mathrm{STA}(t), \rho_\mathrm{STA}(t)]$ and $i \dot{\rho}_0(t)=[H_0(t), \rho_\mathrm{0}(t)]$ for each isentropic stroke. After reaching points $B'$ and $D'$ in Fig.~\ref{fig:fig1}\panel{(b)}, with corresponding states $\rho'_B$ and $\rho'_D$, respectively, the latter get thermalized in the two thermalization strokes until they reach the thermal states $\rho_C$ and $\rho_A$ at points $C$ and $A$, respectively. The initial strengths of the magnetic fields at point $A$ in Fig.~\ref{fig:fig1}\panel{(b)} of the first isentropic stroke are $h_{j,\mathrm{i}}=0.5$ and $b_{j,\mathrm{i}}=0$ for each spin, respectively, with vanishing interactions, $J_{j,\mathrm{i}}=0$. The final magnetic fields at point $B$ in Fig.~\ref{fig:fig1}\panel{(b)} are $h_{j,\mathrm{f}}=0$ and $b_{j,\mathrm{f}}=1$. In order to test the practical applicability of our \emph{local} counter-diabatic term, Eq.~\eqref{eq_Hcd}, we randomly draw $100$ final interaction strengths $J_{j,\mathrm{f}}$ from a Gaussian distribution with standard deviation $\sigma=0.1$ and zero mean. For the isentropic stroke~3 the initial (point $C$) and final (point $D$) parameters are interchanged. The explicit forms of all time-dependent fields are given in Eqs.~\eqref{app_fields_many-body} in Appendix~\ref{sec:A2}). The durations of the thermalization strokes $2$ and $4$ are set to $\tau_2=\tau_4=0.1$, and the cold and hot bath temperatures to $\Tc=0.22$ and $\Th=22$, respectively. 

\par

In order to further improve the shortcut-to-adiabaticity Hamiltonian, Eq.~\eqref{eq:SpinCDHamiltonian}, we numerically optimize the free control parameter $\vartheta_0$ in Eq.~\eqref{eq:lambda}: For each instance of $J_{j,\mathrm{f}}$ and isentropic duration time $\tau$ we determine this optimized parameter via an iterative numerical update until we maximize the success fidelity 
\begin{equation} 
  F(\rho',\rho)\coloneq\Tr\left(\sqrt{\sqrt{\rho'} \rho \sqrt{\rho'}}\right),
  \label{eq:Fidelity}
\end{equation}
where $\rho'$ [$\rho$] are the final states at the end of each isentropic stroke with $H^*_\mathrm{STA}(t)$ [$H_0 (t)$]. We restrict the values of $\vartheta_0$ to be in $[0,1]$ to keep the strengths of the additional magnetic fields $Y_j(\vartheta_0,t)$ at a reasonable level, i.e., not overwhelmingly larger than the other fields in $H_0(t)$. Since the counter-diabatic protocol for the third stroke is the time-reversed version of $H^*_\mathrm{CD}(t)$ for the first stroke we only need to optimize $\vartheta_0$ for the latter and use the same value for the former. All simulations were implemented with QuTip 4.2~\cite{johansson2013qutip}.

\par

Figure~\ref{fig:EfficienciesPowerCounter} shows the power $\mathcal{P}$, work components $W$, efficiency $\eta$, and success fidelity $F$ of our many-body quantum Otto engine (i)~with shortcuts to adiabaticity [Eq.~\eqref{eq:SpinCDHamiltonian}] and (ii) with the original nonadiabatic protocol [Eq.~\eqref{eq:IsingSpinModel}] for different durations $\tau=\tau_1=\tau_3$ of the isentropic strokes. Figure~\ref{fig:EfficienciesPowerCounter}\panel{(a)} reveals that the finite-time cycle under STA always acts as an engine, i.e., it provides useful work (blue area where $\mathcal{P}<0$). The engine still works in the limit $\tau_1,\tau_3 \to 0$, where the cycle time $\tau_\mathrm{cycle}$ is dominated by thermalization. By contrast, the original protocol $H_0(t)$ becomes nonadiabatic and quantum friction impacts its operation. Indeed, for too fast isentropic strokes ($\tau\lesssim 10$) the final states $\rho_B^\prime$ and $\rho_D^\prime$ become so different from their adiabatic counterparts $\rho_B$ and $\rho_D$ that the cycle ceases to describe an engine---rather than delivering power it consumes power.

\par

Figure \ref{fig:EfficienciesPowerCounter}\panel{(b)} presents an operational insight into the engine by depicting the work components $W_0^{1+3}$ [Eq.~\eqref{eq_W0}], attributed to useful work extracted by the piston (load), and $W_\mathrm{CD}^{1+3}$ [Eq.~\eqref{eq_Wcd}], stemming from the external control device. Since the counter-diabatic Hamiltonian is \emph{not} exact a finite amount of work is exchanged between the WM and the control. Up to moderate stroke durations of $\tau\lesssim 5$ a part of the work generated from the heat input is directed into the control and is thus lacking for the piston, i.e., $|W_\mathrm{0}^{1+3}| < |W_\mathrm{STA}^{1+3}|$. In this region [green-shaded area in Fig.~\ref{fig:EfficienciesPowerCounter}\panel{(b)}] the cycle operates as a finite-power heat engine (\cf Sec.~\ref{sec_he}). By contrast, for $5\lesssim\tau\lesssim 30$ (yellow-shaded area), the engine is of a hybrid thermomechanical nature where not only work stemming from the converted heat input but also work stemming from the controller is transferred to the piston, i.e., $|W_\mathrm{0}^{1+3}| > |W_\mathrm{STA}^{1+3}|$ (\cf Sec.~\ref{sec_hybrid}). Finally, the third (gray-shaded) area represents the adiabatic limit where the counter-diabatic term $H^*_\mathrm{CD}(t)$ is very small and thus the entire work $W_\mathrm{STA}^{1+3} \approx W_0^{1+3}$ is performed on the piston. Namely, the work done by the external control device naturally converges towards 0. A more detailed discussion of the work components in the individual strokes is given in Appendix~\ref{app_work_components}.

\par
The efficiency $\eta$ of the engine [Eq.~\eqref{eq_eta_heat}] for the green-shaded and Eq.~\eqref{eq_eta_hybrid} for the yellow-shaded area, respectively) is shown in Fig.~\ref{fig:EfficienciesPowerCounter}\panel{(c)}. As expected, the efficiency of the nonadiabatic engine [governed by the protocol $H_0(t)$ for finite stroke duration] strongly decreases with decreasing stroke duration (red line). By contrast, the STA cycle keeps operating as an engine whose efficiency, albeit being lower than the adiabatic one, is still reasonably high and does not decrease further as the cycle time is further reduced.

\par

Figure~\ref{fig:EfficienciesPowerCounter}\panel{(d)} depicts the success fidelity $F$ [Eq.~\eqref{eq:Fidelity}], i.e., the overlap between the final states $\rho_B'$ and $\rho_B$ for the first isentropic stroke (point $B$) and $\rho_D'$ and $\rho_D$ for the second isentropic stroke (point $D$), respectively. The strong decay of the fidelity $F_1=F(\rho_B',\rho_B)$ after the first isentropic stroke for decreasing isentropic stroke times in the original Otto cycle coincides with the corresponding drops in power and efficiency. It, however, does not decay to 0, as in the quench limit $\tau\to 0$ the state barely changes such that $F(\rho_B',\rho_B) \approx F(\rho_A, \rho_B)$. By contrast, the fidelity $F_3=F(\rho_D',\rho_D)$ of the third stroke remains close to unity for all $\tau$ due to $\Th$ being so high that the eigenstate populations are almost uniform. In the adiabatic limit the fidelity approaches unity, as expected (gray-shaded area).

\par

The situation strongly changes in the presence of the counter-diabatic protocol $H^*_\mathrm{CD}(t)$: Except for a small dip in the thermomechanical-engine regime the fidelities remain close to unity for all times and both isentropic strokes. It is remarkable that this also holds in the limit $\tau\rightarrow 0$ but this limit requires strong magnetic fields for implementing the CD protocol. We further note that the fidelity behaves very similarly for the different instances of the interaction strength. In Appendix~\ref{app_sigma} we show that the fidelity using our \emph{local} method decreases with increasing standard deviation of $J_{j,\mathrm{f}}$. This indicates the need for higher-order counter-diabatic protocols, i.e., the addition of controlled spin interactions rather than only adding local magnetic fields, in such situations.

\par

We expect a positive $\Wcd^{1+3}$ and thus the occurrence of the thermomechanical engine's being an artifact of the \emph{inexact} counter-diabatic drive of the many-body working medium. In Appendix~\ref{App:secA1} we present the above analysis for a single-body quantum Otto cycle where an \emph{exact} counter-diabatic term can be found (see also Ref.~\cite{cakmak2019spin}).

\par
\begin{figure}
  \centering
  \includegraphics[width=.95\columnwidth]{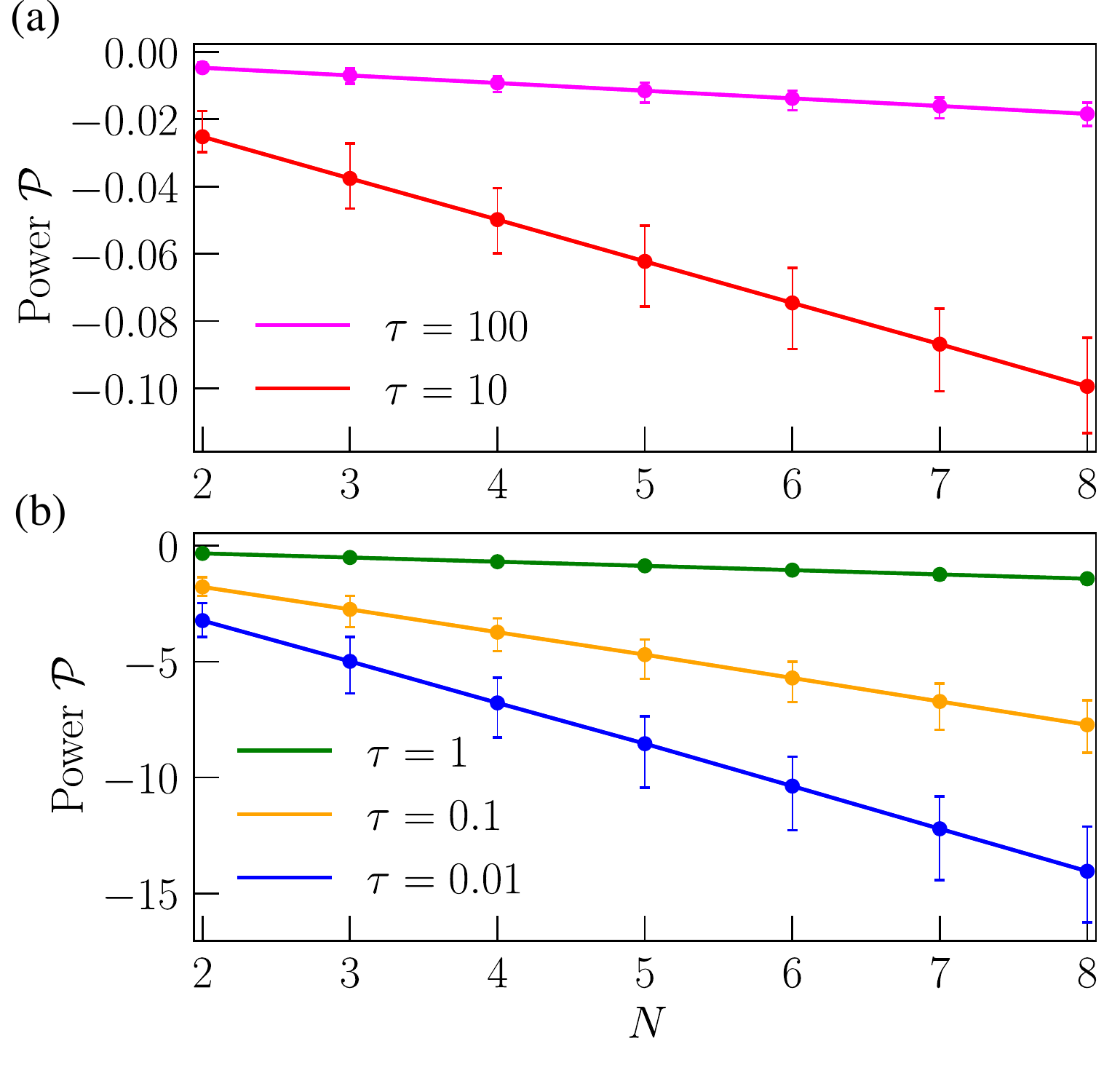}
  \caption{\textbf{Power scaling with the system size.} Power $\mathcal{P}$ [Eq.~\eqref{eq_P_heat}] of the Otto cycle with (a)~the original [Eq.~\eqref{eq:IsingSpinModel}] and (b)~the STA Hamiltonian [Eq.~\eqref{eq:SpinCDHamiltonian}] as a function of the number of spins and for different isentropic stroke times $\tau=\tau_1=\tau_3$. Other parameters are the same as in Fig.~\ref{fig:EfficienciesPowerCounter}.}
  \label{fig:ScalingPowerCD}
\end{figure}
\par

Finally, Fig.~\ref{fig:ScalingPowerCD} shows the power $\mathcal{P}$ [Eq.~\eqref{eq_P_heat}] of our many-body quantum Otto cycle as a function of the system size $N$ for both protocols for different isentropic stroke durations $\tau$. Note that the original protocol operates as an engine only for $\tau\gtrsim 10$, whereas the STA protocol also works for shorter stroke durations [\cf Fig.~\ref{fig:EfficienciesPowerCounter}\panel{(a)}]. It is shown that in either case and independent of $\tau$, the power scales linearly with the number of spins. For future work it would be interesting to consider possible cooperative effects~\cite{manatuly2019collectively}.

\section{Discussion and Outlook}\label{sec_dis}

In this work we have presented a finite-time many-body quantum heat engine with finite power output. It is composed of four strokes (two isentropic and two thermal) and a spin system as its working medium. In its isentropic strokes, work is either extracted from or performed on the WM. The caveat of highly efficient but adiabatic cycles is their requirement for almost infinitely long cycle times, which results in vanishing power (work divided by cycle time). In the context of QHEs the detrimental effect of nonadiabatic (diabatic) evolution on the engine efficiency has been dubbed ``quantum friction''~\cite{kosloff2002discrete,feldmann2003quantum,feldmann2006quantum,kosloff2017quantum}. Its illustrative explanation is that the excitation of coherences costs an extra amount of energy but these coherences are ``dissipated away'' in the subsequent thermalization strokes in which the working medium is put into contact with a thermal bath. Hence, while in classical heat engines the non-reversibility of adiabatic strokes due to entropy-increasing friction causes the engine efficiency to drop, quantum-mechanically, the quantum nonadiabatic behavior may create quantum friction devoid of any change in the WM entropy. Mathematically, quantum friction can only occur if the control Hamiltonian does not commute with itself at different times~\cite{kosloff2017quantum}.

\par 

A possible solution to overcome this problem of zero power output at finite cycle times is so-called shortcuts to adiabaticity (STA), which have been developed in the context of adiabatic quantum computation and have later also been applied in the field of quantum thermodynamics~\cite{deng2013boosting,deffner2014classical,delcampo2014more,beau2016scaling,abah2017energy,campbell2017tradeoff,patra2017shortcuts,abah2018performance,diao2018shortcuts,duncan2018shortcuts,alipour2019shortcuts,abah2019shortcut,cakmak2019spin,delcampo2019focus,bason2012high,an2016shortcuts}. A major obstacle has been to find an easy-to-implement STA method for many-body systems since the additional counter-diabatic term normally requires \emph{a priori knowledge} of the system eigenstates at all times. In this work we employ local CD driving~\cite{sels2017minimizing,hartmann2019rapid} to a many-body QHE where the additional CD drive consists of adding a local magnetic field in the $y$ direction to speed up the engine while minimizing the quantum friction. The latter is further reduced by an iterative variation of the free control parameter $\vartheta_0$ for the magnetic-field strength.

\par 

To assess the performance of the many-body QHE we compare (i) the engine with the STA protocol, i.e, $H^*_\mathrm{STA}(t)$, and (ii) the engine without the STA protocol, i.e., $H_0(t)$. The sped-up QHE with STA shows large improvement in power output, efficiency, and success fidelity for finite times compared to the original, nonadiabatic QHE without STA, in particular, for short isentropic stroke durations where the QHE without STA ceases to work as an engine. However, for such short cycle times the additional magnetic field required by the STA may be much larger than the other fields. Namely, the dynamics of the working medium may then become dominated by the CD protocol $H^*_\mathrm{CD}(t)$.

\par 

As our additional CD term is \emph{not} exact, we have to take care of the energetic balance of the external controller that implements $H^*_\mathrm{CD}(t)$. In particular, we have to distinguish cases where the controller receives work from or provides work to the engine, respectively: If the controller \emph{receives} the work $\Wcd^{1+3}$, the QHE works as a genuine heat engine where only thermal energy is converted into mechanical work. By contrast, if the controller \emph{provides} the work $\Wcd^{1+3}$, the QHE works as a hybrid heat engine where thermal as well as mechanical work is converted into mechanical work. As we aim for a sped-up engine powered by heat rather than by an external battery, we want to avoid the latter case. The additional input work can be seen as an artifact of the \emph{inexact} counter-diabatic drive and thus shows the importance of striving for an \emph{exact} CD drive where the useful work is only done on the piston. This exact drive, however, may be very challenging to implement experimentally, as it may involve controlled many-body interactions rather than simply applying \emph{local} additional magnetic fields on each spin. We note that this trade-off between the exactness of the protocol and its experimental realizability naturally occurs in the many-body case. By contrast, in the single-body case, exact and conceptually simple protocols can often be found. Note further, that these \emph{operational} costs conceptually differ from the costs of \emph{implementing} the CD Hamiltonian, for which different quantifiers have been suggested in the literature~\cite{abah2017energy,campbell2017tradeoff,zheng2016cost, abah2019shortcut, abah2018performance,cakmak2019spin,tobalina2019vanishing, abah2019shortcutto}. Incorporating the latter, the efficiency of our sped-up QHE at very short cycle times may decrease considerably. Implementation costs are highly system dependent and may thus be difficult to assess. By contrast, our operational approach is motivated by an engineer that (i) wants to measure the works $W_0$ and $W_{\mathrm{CD}}$, respectively, and (ii)~can distinguish between work output and heat input, which is an intuitive way to define the efficiency of a heat engine.

\par

For future research, we aim at proposing a sped-up many-body quantum Otto engine using an experimentally-feasible lattice gauge architecture~\cite{lechner2015quantum} and to apply our local counter-diabatic method to open many-body quantum systems. A further topic of interest is many-body quantum refrigerators, which may be sped up in analogy to the engines in this work.

\section*{Acknowledgments}

W.\,N.\ acknowledges support from an ESQ fellowship of the Austrian Academy of Sciences (\"OAW). V.\,M.\ acknowledges support from a Start-up Research Grant (Project No.\ SRG/2019/000411) and from a Seed Grant from IISER Berhampur. The research was funded by the Austrian Science Fund (FWF) through a START grant under Project No.~Y1067-N27 and SFB BeyondC Project No.~F7108-N38, the Hauser-Raspe Foundation, and the European Union's Horizon 2020 research and innovation program under Grant Agreement No.~817482 PasQuanS.

\appendix

\section{Single-body quantum working medium}\label{App:secA1}

The many-body quantum Otto cycle presented in the text extends the ideas of a shortcut-to-adiabaticity quantum Otto cycle with a single-body working medium which we consider in this section (see also Ref.~\cite{cakmak2019spin}). To this end we consider the single-spin Landau--Zener (LZ) model with Hamiltonian
\begin{equation}
  H_{\mathrm{LZ},0}(t)=-h_x(t) \sigma^x-b_z(t) \sigma^z.
\label{eq:LZ}
\end{equation}
For the explicit forms of the magnetic fields $h_x(t)$ and $b_z(t)$ we chose
\begin{subequations}
  \begin{align}
    h_x(t)&=h_{x,\mathrm{i}}+(h_{x,\mathrm{f}}-h_{x,\mathrm{i}})\sin^2\left[\dfrac{\pi}{2}\sin^2\left(\dfrac{\pi t}{2 \tau}\right)\right],  \\
    b_z(t)&=b_{z,\mathrm{i}}+(b_{z,\mathrm{f}}-b_{z,\mathrm{i}})\sin^2\left[\dfrac{\pi}{2}\sin^2\left(\dfrac{\pi t}{2 \tau}\right)\right].
  \end{align}
\end{subequations}

\par
\begin{figure}
  \centering
  \includegraphics[width=0.95\columnwidth]{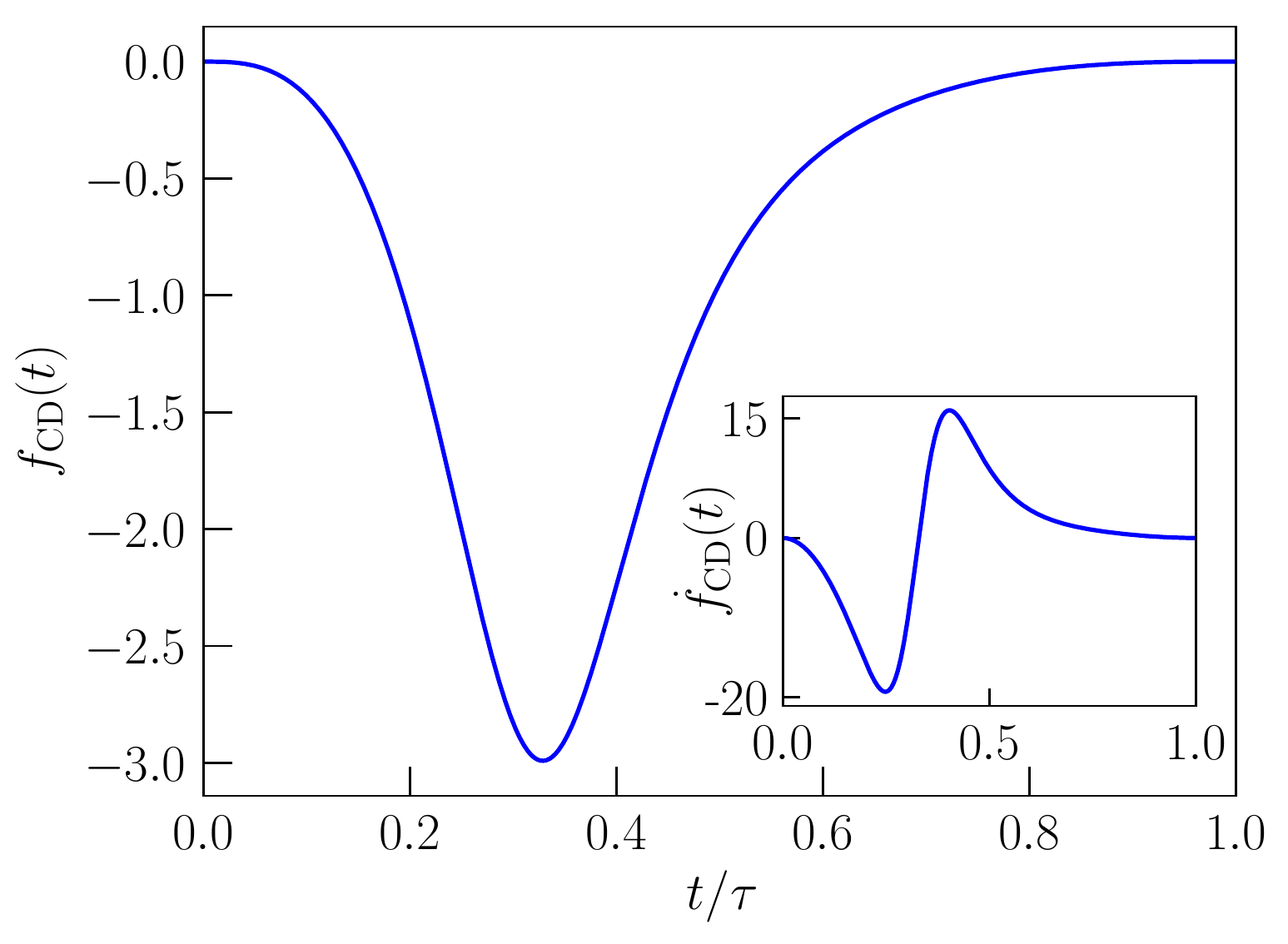}
  \caption{\textbf{Counter-diabatic function.} The function $f_{\mathrm{CD}}(t)$ [Eq.~\eqref{eq:fcd}] of the counter-diabatic protocol, Eq.~\eqref{eq:additionalCD-LZ}, and its time derivative $\dot{f}_{\mathrm{CD}}(t)$ (inset) for isentropic-stroke duration $\tau$. Parameters: $h_{x,\mathrm{i}}=0.1$, $b_{z,\mathrm{i}}=0$, $h_{x,\mathrm{f}}=0$, and $b_{z,\mathrm{f}}=0.5$.}
  \label{fig:CD-LZ}
\end{figure}
\par

The \emph{exact} counter-diabatic term for Hamiltonian~\eqref{eq:LZ} reads~\cite{takahashi2013transitionless,cakmak2019spin}
\begin{equation}
  H_{\mathrm{LZ,CD}}(t)=f_\mathrm{CD}(t)\sigma^y,
\label{eq:additionalCD-LZ}
\end{equation}
where the function
\begin{equation}
  f_{\mathrm{CD}}(t)=\dfrac{1}{2}\dfrac{\dot{h}_x(t) b_z(t) - \dot{b}_z(t) h_x(t)}{h_x^2(t) +b_z^2(t)}
\label{eq:fcd}
\end{equation}
is shown in Fig.~\ref{fig:CD-LZ}.

\par

Thus, the shortcut-to-adiabaticity Hamiltonian reads
\begin{align}
  &H_{\mathrm{LZ,STA}}(t)=H_{\mathrm{LZ},0}(t)+H_{\mathrm{LZ,CD}}(t) \nonumber \\
  &=-h_x(t) \sigma^x-b_z(t) \sigma^z+\dfrac{1}{2}\dfrac{\dot{h}_{x}(t) b_z(t) - \dot{b}_z(t) h_x(t)}{h_x^2(t) +b_z^2(t)} \sigma^y.
    \label{eq:CD-LZ}
\end{align}

\par
\begin{figure*}
  \centering
  \includegraphics[width=0.97\textwidth]{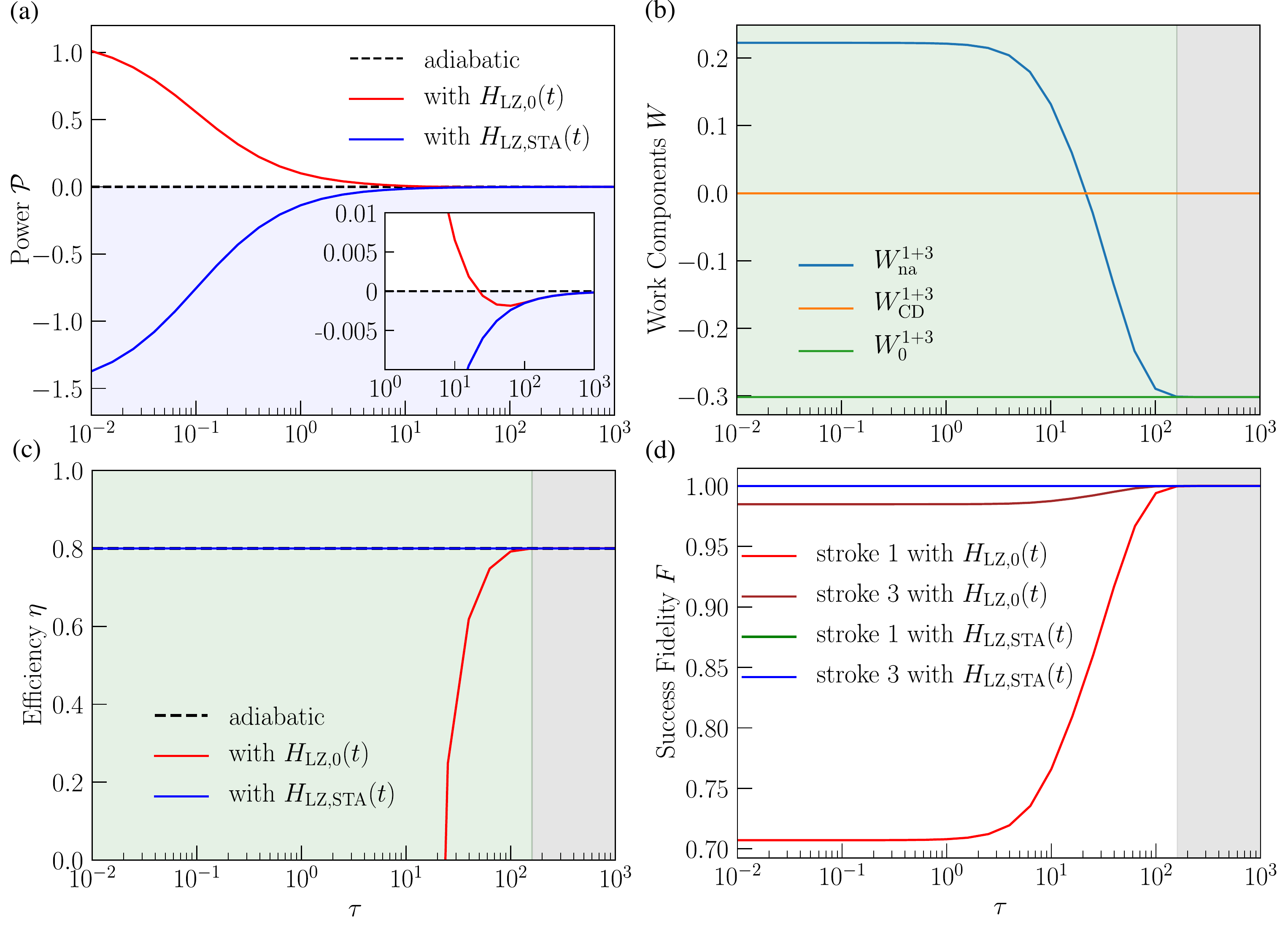}
  \caption{\textbf{Numerical simulation of the finite-time quantum heat engine with a single-body working medium.} (a)~Power $\mathcal{P}$ [Eq.~\eqref{eq_P_heat}] of the Otto cycle whose working medium is governed by (i)~the original protocol $H_{\mathrm{LZ},0}(t)$ [Eq.~\eqref{eq:LZ}] and (ii)~the shortcut-to-adiabaticity protocol $H_\mathrm{LZ,STA}(t)$ [Eq.~\eqref{eq:CD-LZ}] as a function of the isentropic-stroke duration $\tau=\tau_1=\tau_3$. Inset: Zoom-in on the region where the original protocol ceases to describe an engine. (b)~Work components $W$ of the QHE with both Hamiltonians. Since the counter-diabatic drive is \emph{exact} the work $\Wcd$ stemming from the external control device is 0. (c)~Efficiency $\eta$ [Eq.~\eqref{eq_eta_heat}]. Since the counter-diabatic drive is exact, the efficiency of the sped-up cycle equals the efficiency of the adiabatic Otto engine. (d)~Success fidelity [Eq.~\eqref{eq:Fidelity}] of the STA protocol, i.e., the overlap of the states $\rho_{B'}$ and $\rho_{D'}$ with the ideal states $\rho_B$ and $\rho_D$, respectively. Parameters: $\tau_1=\tau_3=\tau$, $\tau_2=\tau_4=0.1$, $\Tc=0.02$, $\Th=2$, $h_{x,\mathrm{i}}=0.1$, $b_{z,\mathrm{i}}=0$, $h_{x,\mathrm{f}}=0$, and $b_{z,\mathrm{f}}=0.5$.}
  \label{fig:figEfficiency}
\end{figure*}
\par

In analogy to Fig.~\ref{fig:EfficienciesPowerCounter} in Sec.~\ref{sec_pfn} of the text, Fig.~\ref{fig:figEfficiency} shows the power $\mathcal{P}$, work components $W$, efficiency $\eta$ and success fidelity $F$ for this single-body quantum Otto engine (i) with shortcuts to adiabaticity [Eq.~\eqref{eq:CD-LZ}] and (ii) with the original nonadiabatic protocol [Eq.~\eqref{eq:LZ}] for different isentropic-stroke durations $\tau=\tau_1=\tau_3$.

\par

Figure~\ref{fig:figEfficiency}\panel{(a)} shows that our STA cycle always works as an engine, i.e., it provides useful work (blue-shaded area) also for short cycle times. By contrast, the Otto cycle governed by the original protocol $H_0(t)$ is hampered by quantum friction and ceases to work as an engine for $\tau\lesssim 25$. As for the many-body engine in Fig.~\ref{fig:EfficienciesPowerCounter}\panel{(a)}, for too short cycle times the final states $\rho_B'$ and $\rho_D'$ are so different from the adiabatic states $\rho_B$ and $\rho_D$ that the cycle consumes rather than delivers power.

\par

Figure~\ref{fig:figEfficiency}\panel{(b)} depicts the work components $W_0^{1+3}$ [Eq.~\eqref{eq_W0}] attributed to useful work extracted by the load and $\Wcd^{1+3}$ [Eq.~\eqref{eq_Wcd}] stemming from the external control device. Since the counter-diabatic term is \emph{exact}, the work $\Wcd^{1+3}$ stemming from the external control device is 0 and thus the entire work $W_\mathrm{STA}^{1+3}$ produced by the engine is performed on the piston, i.e., $\Wsta^{1+3}=W_0^{1+3}$. Namely, the external controller optimally assists the piston (green-shaded area). The gray-shaded area corresponds to the adiabatic limit.

\par

Figure~\ref{fig:EfficienciesPowerCounter}\panel{(c)} shows the efficiency $\eta$ [Eq.~\eqref{eq_eta_heat}] of the single-body Otto cycle. Since the counter-diabatic drive is exact, the work per cycle with STA equals the work per cycle in the adiabatic engine. Hence, the efficiency of the STA engine equals the efficiency of the adiabatic engine.

\par

Finally, Fig.~\ref{fig:EfficienciesPowerCounter}\panel{(d)} depicts the success fidelity $F$ [Eq.~\eqref{eq:Fidelity}] as the overlap between the final states $\rho_B'$ and $\rho_B$ for the first isentropic stroke [point $B$ in Fig.~\ref{fig:fig1}\panel{(b)}] and $\rho_D'$ and $\rho_D$ for the isentropic stroke 3 (point $D$), respectively. As the counter-diabatic term is exact, the corresponding fidelity is always unity. By contrast, as in Fig.~\ref{fig:EfficienciesPowerCounter}\panel{(d)} the fidelities in the original strokes decrease with decreasing cycle times and converge towards their quench values $F(\rho_A,\rho_B)$ and $F(\rho_C,\rho_D)$, respectively.

\section{Many-body quantum working medium}\label{sec:A2}

\subsection{Shortcut-to-adiabaticity protocols}\label{app_sta_protocols}

For the original [Eq.~\eqref{eq:IsingSpinModel}] and shortcut-to-adiabaticity Hamiltonian [Eq.~\eqref{eq:SpinCDHamiltonian}] of the many-body WM in the text we chose the following time dependence of the magnetic fields and interaction strengths,
\begin{subequations}\label{app_fields_many-body}
  \begin{align}
    h_j(t)&=h_{j,\mathrm{i}}+(h_{j,\mathrm{f}}-h_{j,\mathrm{i}})\sin^2\left[\dfrac{\pi}{2}\sin^2\left(\dfrac{\pi t}{2 \tau}\right)\right]  \\
    b_j(t)&=b_{j,\mathrm{i}}+(b_{j,\mathrm{f}}-b_{j,\mathrm{i}})\sin^2\left[\dfrac{\pi}{2}\sin^2\left(\dfrac{\pi t}{2 \tau}\right)\right]  \\
    J_j(t)&=J_{j,\mathrm{i}}+(J_{j,\mathrm{f}}-J_{j, \mathrm{i}})\sin^2\left[\dfrac{\pi}{2}\sin^2\left(\dfrac{\pi t}{2 \tau}\right)\right].\label{eq_app_J}
  \end{align}
\end{subequations}
The time derivative of the control function, Eq.~\eqref{eq:lambda}, is
\begin{equation}
  \dot{\vartheta}(\vartheta_0,t)=\vartheta_0\dfrac{\pi^2}{4\tau}\sin\left(\dfrac{\pi}{\tau}t\right)\sin\left[\pi \sin^2 \left(\dfrac{\pi}{2\tau}t \right)\right],
\end{equation}
where $\vartheta_0$ is the free control parameter that is iteratively optimized. Note that the chosen function, Eq.~\eqref{eq:lambda}, assures smoothness, i.e., $\dot{\vartheta}(t=0)=\dot{\vartheta}(t=\tau_j)=\ddot{\vartheta}(t=0)=\ddot{\vartheta}(t=\tau_j)=0$, in the beginning and end of the isentropic strokes $j \in \{1,3\}$.

\subsection{Work components of the isentropic strokes}\label{app_work_components}

\par
\begin{figure}
  \centering
  \includegraphics[width=0.95\columnwidth]{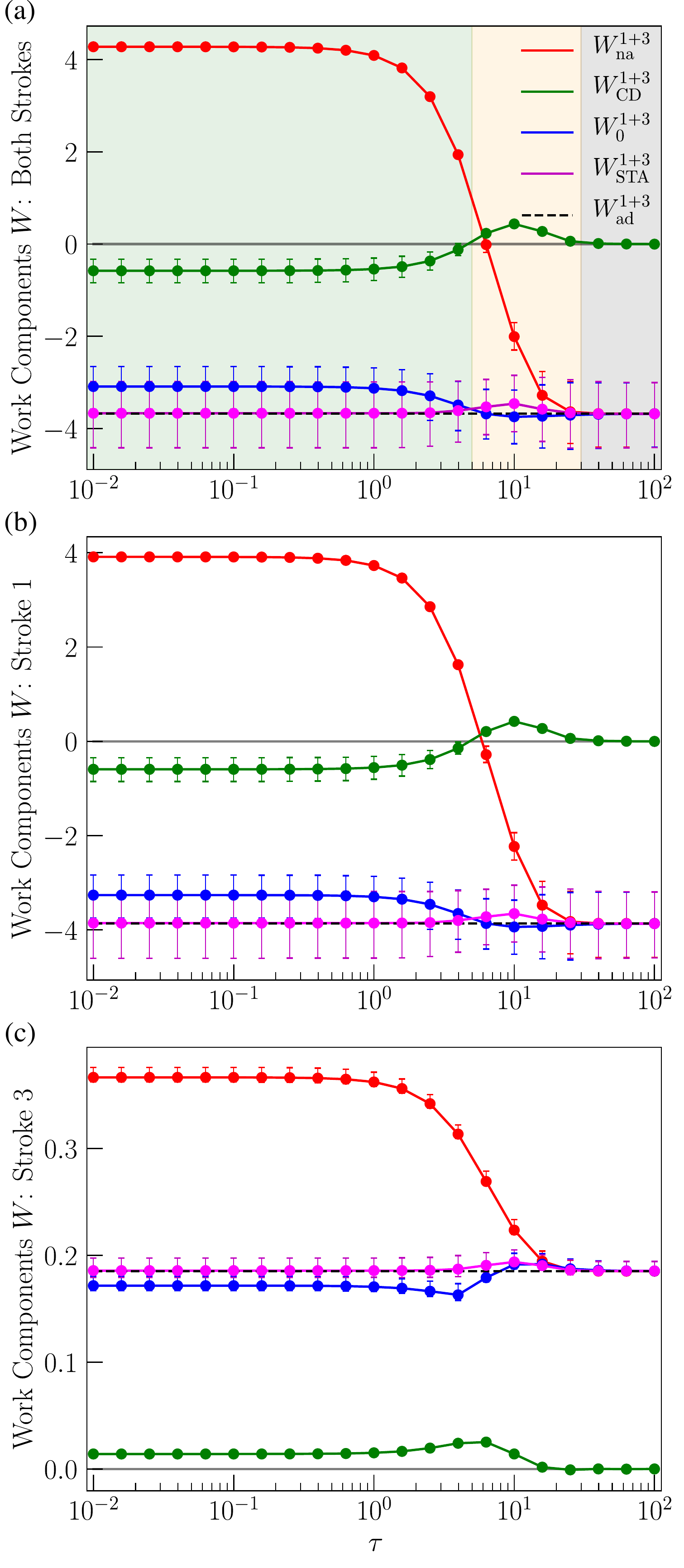}
  \caption{\textbf{Work components of the finite-time quantum heat engine with a many-body working medium.} Work components $W_0^{1,3}$, stemming from the piston, and $W_\mathrm{CD}^{1,3}$, stemming from the external control device, for the quantum Otto cycle with (i)~the shortcut-to-adiabaticity Hamiltonian [Eq.~\eqref{eq:SpinCDHamiltonian}] and (ii)~the original Hamiltonian [Eq.~\eqref{eq:IsingSpinModel}] (red curve). The total works in~(a) are the sum of the respective works in stroke~$1$~(b) and stroke~$3$~(c), respectively. Same parameters as in Fig.~\ref{fig:EfficienciesPowerCounter}.} 
  \label{fig:work}
\end{figure}
\par

Figure~\ref{fig:work}\panel{(a)} shows the work components $W_0^{1+3}$ [Eq.~\eqref{eq_W0}] and $W_{\mathrm{CD}}^{1+3}$ [Eq.~\eqref{eq_Wcd}] of the total work $W_\mathrm{STA}^{1+3}$ [Eq.~\eqref{eq_Wsta}] over a cycle [\cf Fig.~\ref{fig:EfficienciesPowerCounter}\panel{(b)}]. Additionally, the work $\Wna^{1+3}$ of the original nonadiabatic finite-time cycle is shown (red curve). These total works are the sum of the individual work components in strokes $1$ and $3$, shown in Figs.~\ref{fig:work}\panel{(b)} and~\ref{fig:work}\panel{(c)}, respectively.

\subsection{Effect of the different  interaction strengths}\label{app_sigma}

\par
\begin{figure}
  \centering
  \includegraphics[width=0.95\columnwidth]{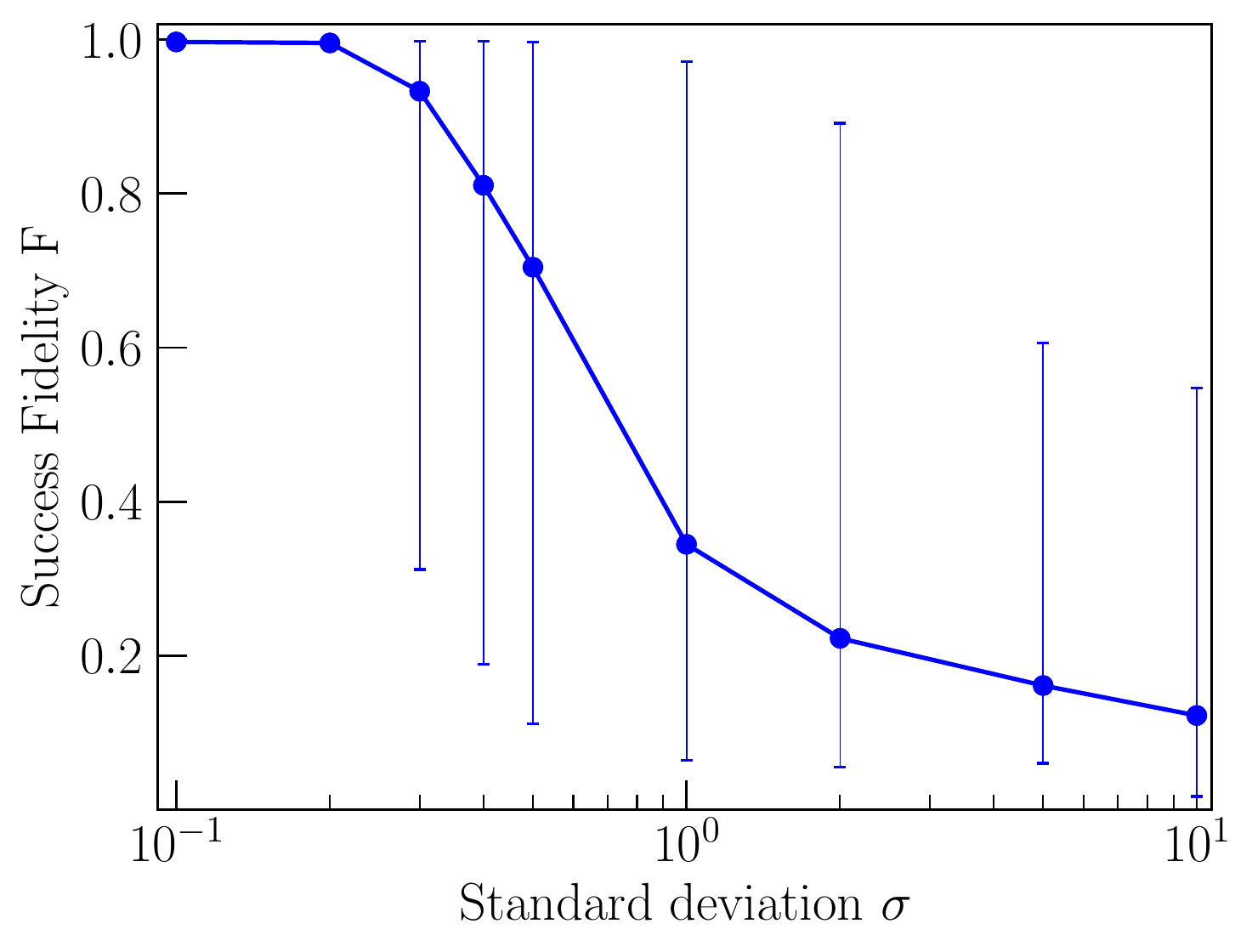}
  \caption{\textbf{Success fidelities for different standard deviations.}
    Success fidelity $F_1=F(\rho_B', \rho_B)$ [Eq.~\eqref{eq:Fidelity}] for the STA Hamiltonian [Eq.~\eqref{eq:SpinCDHamiltonian}] for a fixed stroke duration $\tau_1=0.1$ as a function of the standard deviation $\sigma$ of the final interaction strengths $J_{j,\mathrm{f}}$ (see Sec.~\ref{sec_pfn}). Other parameters are the same as in Fig.~\ref{fig:EfficienciesPowerCounter}.} 
  \label{fig:sigma}
\end{figure}
\par

Our STA protocol, Eq.~\eqref{eq:SpinCDHamiltonian}, being of a local nature, the question of how good this local approximation works for the operation of a sped-up quantum Otto engine naturally arises. In Sec.~\ref{sec_pfn} we have introduced different final interaction strengths $J_{j,\mathrm{f}}$ drawn from a Gaussian distribution with zero mean and standard deviation $\sigma=0.1$. As shown by the distribution of power, work, efficiency, and success fidelity in Fig.~\ref{fig:EfficienciesPowerCounter}, for this value of $\sigma$ our ansatz performs reasonably well.

\par

As a variation of Fig.~\ref{fig:EfficienciesPowerCounter}\panel{(d)}, we have investigated the impact of larger standard deviations of the final interaction strengths on the success fidelity of the first stroke for a fixed stroke duration of $\tau_1=0.1$ (Fig.~\ref{fig:sigma}). It is shown that on average the larger $\sigma$, the smaller the fidelity. This indicates the need for higher-order counter-diabatic protocols, i.e., the addition of controlled spin interactions, rather than only local magnetic fields, in such situations.

\section{Approximate adiabatic gauge potential}\label{sec:AppendixB}

Here we describe in detail the derivation of the adiabatic gauge potential following the method in Ref.~\cite{sels2017minimizing}. For the sake of readability we mainly omit explicit time dependences throughout this section.

\par
The Hamiltonian $H_0$ in the rotating frame with respect to a unitary $U(\vartheta(t))$ has the form
\begin{equation}
\tilde{H}_\mathrm{m}=\tilde{H}_0-\dot{\vartheta}\tilde{\mathcal{A}}_{\vartheta}, \label{eq:eq3}
\end{equation}
where $\tilde{H}_0=U^{\dagger} H_0U$ is the diagonalized (stationary) instantaneous Hamiltonian and $\tilde{\mathcal{A}}_{\vartheta}$ the adiabatic gauge potential in the rotating frame with respect to the time-dependent variable $\vartheta$ describing the dynamics of the system. The Hamiltonian $\tilde{H}_0$ is diagonal and thus all diabatic transitions occur due to the adiabatic gauge potential in the second term in Eq.~\eqref{eq:eq3}.

\par

The form of the Hamiltonian in the moving frame, i.e., Eq.~\eqref{eq:eq3}, can be derived by evolving a quantum state $\ket{\psi}$ according to the Schr\"odinger equation $i \hbar \partial_t \ket{\psi} = H_0(\vartheta(t)) \ket{\psi} $ with a time-dependent Hamiltonian $H_0(\vartheta(t))$ in a rotating frame $\ket{\tilde{\psi}}=U^{\dagger} \ket{\psi}$. This leads to
\begin{align}
\tilde{H}_\mathrm{m} \ket{\tilde{\psi}}&= i \hbar \partial_\vartheta \ket{\tilde{\psi}} =i \hbar \partial_\vartheta (U^{\dagger} \ket{\psi}) \nonumber \\
&=i \hbar \partial_\vartheta U^{\dagger} \ket{\psi}+i \hbar U^{\dagger} \partial_\vartheta \ket{\psi} \nonumber \\
&=i \hbar \partial_t \vartheta \, \partial_{\vartheta} U^{\dagger} \ket{\psi}  + U^{\dagger} H_0 \ket{\psi}  \nonumber \\
&=\partial_t \vartheta \, (i \hbar \partial_{\vartheta} U^{\dagger} U) \ket{\tilde{\psi}} + U^{\dagger} H_0 U \ket{\tilde{\psi}} \nonumber \\ & = (\tilde{H}_0 - \dot{\vartheta} \tilde{\mathcal{A}}_{\vartheta}) \ket{\tilde{\psi}}, \label{eq:eqA1}
\end{align}
with the adiabatic gauge potential
\begin{equation}
\mathcal{A}_{\vartheta}\coloneq- i \hbar (\partial_{\vartheta} U^{\dagger})U = i \hbar U^{\dagger} \partial_{\vartheta}U. \label{eq:eqA2}
\end{equation}
Differentiating $\tilde{H}_0(\vartheta)=U^{\dagger}(\vartheta) H_0(\vartheta) U(\vartheta)$ with respect to the system's dynamical parameter $\vartheta$, we obtain
\begin{align}
\partial_{\vartheta}\tilde{H}_0=U^{\dagger} \partial_{\vartheta} H_0 U+\dfrac{i}{\hbar}[\tilde{\mathcal{A}}_{\vartheta},\tilde{H}_0]. \label{eq:eqA3}
\end{align}
Transforming back to the laboratory frame and using that the gauge potential eliminates the off-diagonal terms of $\tilde{H}_\mathrm{m}$, i.e., $[\partial_{\vartheta} \tilde{H}_0, \tilde{H}_0]=0$, we obtain
\begin{equation}
[i \partial_{\vartheta} H_0-[\mathcal{A}_{\vartheta},H_0], H_0]=0.
\label{eq:eq4}
\end{equation}
The solution $\mathcal{A}_\vartheta$ of this equation gives the \emph{exact} counter-diabatic Hamiltonian $\Hcd(t) = \dot{\vartheta}(t) \mathcal{A}_\vartheta(t)$ in Eq.~\eqref{eq:eq2} (see Ref.~\cite{kolodrubetz2017geometry} for more details). In the instantaneous eigenbasis, it reads~\cite{berry2009transitionless}
\begin{align}
\Hcd(t)&=i \hbar \sum_n \left(\ketbra{\partial_t n}{n}-\langle n|\partial_t n \rangle \proj{n}\right) \nonumber \\
& =i \hbar \sum_{m \neq n} \sum_n  \dfrac{\proj{m}\partial_t H_0 \proj{n}}{E_m-E_n}.
\label{eq:CDHamiltonian}
\end{align}

\par

Equation~\eqref{eq:CDHamiltonian} requires \emph{a priori} knowledge of all eigenstates at all times during the sweep and is therefore impracticable, especially in a many-body setup. Hence, we strive for an approximate solution $\mathcal{A}^*_\vartheta$ for the adiabatic gauge potential that can relatively easily be implemented in experiments. To this end we employ the variational principle method of Ref.~\cite{sels2017minimizing}, namely, that solving Eq.~\eqref{eq:eq4} is equivalent to minimizing the Hilbert-Schmidt norm of the Hermitian operator
\begin{equation}
G_{\vartheta}(\mathcal{A}^*_{\vartheta})=\partial_{\vartheta} H_0+i[\mathcal{A}^*_{\vartheta},H_0] \label{eq:eq5}
\end{equation}
with respect to the parameters of $\mathcal{A}^*_{\vartheta}$ and $\hbar \equiv 1$. Here, we seek the minimum of the operator distance
\begin{equation}
  D^2(\mathcal{A}^*_{\vartheta})=\Tr\left\{\left[G_{\vartheta}(\mathcal{A}_{\vartheta})-G_{\vartheta}(\mathcal{A}^*_{\vartheta})\right]^2\right\}
\end{equation}
between the exact $G_{\vartheta}(\mathcal{A}_{\vartheta})$ and the approximate $G_{\vartheta}(\mathcal{A}^*_{\vartheta})$. Minimizing this operator distance is equivalent to minimizing the action 
\begin{equation}
\mathcal{S}(\mathcal{A}^*_{\vartheta})=\Tr[G^2_{\vartheta}(\mathcal{A}^*_{\vartheta})] \label{eq:eq6}
\end{equation}
associated with the parameters of the approximate adiabatic gauge potential $\mathcal{A}^*_{\vartheta}$, i.e.,
\begin{equation}
\dfrac{\delta \mathcal{S}(\mathcal{A}^*_{\vartheta})}{\delta \mathcal{A}^*_{\vartheta}}=0, \label{eq:eq7}
\end{equation}
where $\delta$ denotes the functional derivative (see Refs.~\cite{sels2017minimizing} and~\cite{kolodrubetz2017geometry} for more details). 

\section{Approximate gauge potential for the Ising spin model}\label{sec:AppendixB1}

For the Ising spin Hamiltonian, Eq.~\eqref{eq:IsingSpinModel}, from the text and the ansatz, Eq.~\eqref{eq:ansatz1}, the operator $G_t(\mathcal{A}^*_t) = \dot{\vartheta}(t) G_\vartheta(\mathcal{A}^*_\vartheta)$ with $G_\vartheta(\mathcal{A}^*_\vartheta)$ according to Eq.~\eqref{eq:eq5} reads
\begin{align}
G_t(\mathcal{A}^*_t)&=-\sum_{j=1}^N(\dot{h}_j-2 \alpha_j b_j)\sigma_j^x+(\dot{b}_j+2 \alpha_j h_j)\sigma_j^z \nonumber \\
&+\sum_{j=1}^{N}2 \alpha_j (J_j \sigma_j^x\sigma_{j+1}^z +J_{j-1} \sigma_{j-1}^z \sigma_{j}^x)-\dot{J}_j\sigma_j^z\sigma_{j+1}^z
\label{Appendix:HermitianIsing}
\end{align}
and the action, Eq.~\eqref{eq:eq6}, with respect to time evaluates to
\begin{multline}
\dfrac{\mathcal{S}(\alpha_j)}{2^N N}=(\dot{h}_j-2 \alpha_j b_j)^2+(\dot{b}_j+2 \alpha_j h_j)^2 \\
+(\dot{J}_j)^2+4 \alpha_j^2 (J_{j-1}^2+J_j^2),
\end{multline}
where $2^N$ is the dimension of the Hilbert space and the $N$ in the denominator stems from the summation over $N$ identical instances. Minimizing this action with respect to $\alpha_j$ then yields Eq.~\eqref{eq:alphaspin} in the text.

\end{document}